\documentclass[aps,nofootinbib,notitlepage,superscriptaddress,twocolumn,10pt,prd]{revtex4-2}

\usepackage{bm}
\usepackage{graphicx}
\usepackage{amsmath,amssymb}
\usepackage{hyperref}
\usepackage{braket}
\usepackage{subfigure}
\usepackage{float}
\usepackage[dvipsnames]{xcolor}
\usepackage{soul}
\usepackage{rotating}
\usepackage{multirow}
\usepackage{mathtools}

\usepackage{makecell}

\usepackage[margin=0.75in]{geometry}

\hypersetup{
	colorlinks=true,
	linkcolor=red,
	citecolor=blue,
}
\usepackage{physics}

\usepackage{natbib}
\usepackage{verbatim}
\usepackage{dcolumn}
\usepackage[amssymb]{SIunits}
\usepackage{tabularx}
\usepackage{booktabs}
\usepackage[normalem]{ulem}

\newcommand{\be}{\begin{equation}}
\newcommand{\ee}{\end{equation}}
\newcommand{\ba}{\begin{eqnarray}}
\newcommand{\ea}{\end{eqnarray}}

\definecolor{ForestGreen}{RGB}{36,179,0}

\begin{document}

\title{Reconstructing Quintessence}

\author{Minsu Park}
\affiliation{Center for Particle Cosmology, Department of Physics and Astronomy, University of Pennsylvania, Philadelphia, PA 19104, USA}
\author{Marco Raveri}
\affiliation{Center for Particle Cosmology, Department of Physics and Astronomy, University of Pennsylvania, Philadelphia, PA 19104, USA}
\author{Bhuvnesh Jain}
\affiliation{Center for Particle Cosmology, Department of Physics and Astronomy, University of Pennsylvania, Philadelphia, PA 19104, USA}

\begin{abstract}
We present an Effective Field Theory based reconstruction of quintessence models of dark energy directly from cosmological data. 
We show that current cosmological data possess enough constraining power to test  several quintessence model properties for redshifts $z\in [0,1.5]$ with no assumptions about the behavior of the scalar field potential. We use measurements of the cosmic microwave background, supernovae distances, and the clustering and lensing of galaxies to constrain the evolution of the dark energy equation of state, Swampland Conjectures, the shape of the scalar field reconstructed potential, and the structure of its phase space. 
The standard cosmological model still remains favored by data and, within quintessence models, deviations from its expansion history are bounded to be below the 10\% level at 95\% confidence at any redshift below $z=1.5$.
\end{abstract}

\maketitle

\section{Introduction} \label{Sec:Intro}
Despite the successes of the cosmological constant plus cold dark matter ($\Lambda$CDM) model at explaining a wide range of cosmological observations, the physical origin of the driver of cosmic acceleration remains still unknown.
The $\Lambda$CDM model assumes a constant dark energy (DE) density, with equation of state of $w_{\rm DE}=-1$. 
This is not the only possible driver of cosmic acceleration, and several others have been proposed (see~\cite{Clifton:2011jh, Silvestri:2009hh, Joyce:2014kja} for  reviews).

Quintessence~\cite{Ratra:1987rm} is the simplest of these models.
It consists of a single scalar field with a canonical kinetic term, a potential that distinguishes different models, and no non-gravitational interactions with other sectors.
Quintessence models are able to mimic the effect of the cosmological constant, while also possessing enough degrees of freedom to open up a wide range of theoretical and phenomenological possibilities that can be observed and tested with cosmological data. 
These considerations make quintessence an interesting prototype for deviations from the $\Lambda$CDM model and have spawned a wide array of theoretical studies of its cosmology (see~\cite{Tsujikawa:2013fta} for a review).

These studies range from dynamical analyses to find attractor solutions~\cite{Copeland:1997et} to the classification of different families of potentials into, for example, ``thawing'' and ``freezing'' depending on their cosmological behavior~\cite{Caldwell:2005tm}.
Quintessence models are also at the hearth of the Swampland Conjectures which put bounds on their properties, based on string theoretic considerations of cosmic expansion~\cite{Agrawal:2018own, Ooguri:2018wrx}.

Much of these studies depend on specific models for the quintessence potential or specific model assumptions. These models have desirable qualities from a theoretical perspective, such as single scale dependence and existence of attractor solutions.
Here we take a data first approach to reconstruct the cosmological behavior of quintessence and examine what the data has to say about the various aspects of the model. In particular, we do not assume the monotonicity of the potential, any closed form for the potential, or the existence of an attractor.
Exploiting the Effective Field Theory of Dark Energy~\cite{Gubitosi:2012hu, Bloomfield:2012ff} (see~\cite{Frusciante:2019xia} for a review) and following the strategy outlined in~\cite{Raveri:2019mxg}, we reconstruct the family of quintessence models permitted by the data.
We use a comprehensive array of cosmological observations to constrain different aspects of quintessence models that have been previously discussed only theoretically.
In doing so, we also investigate what each piece of cosmological observable tells us, what the limiting factors are in our ability to constrain such theories, and what future data may be able to tell us about these theories.

The paper is organized as follows. In Sec.~\ref{Sec:Theory} we briefly review quintessence models and the model properties that we will be reconstructing from data.
In Sec.~\ref{Sec:Methods} we review the data reconstruction methods and the datasets used for this work. 
We present our results in full in Sec.~\ref{Sec:Results} and we conclude with a summary of our results and discussion in Sec.~\ref{Sec:Conclusions}. 

\section{Quintessence reconstruction} \label{Sec:Theory}

The quintessence model of dark energy consists of a scalar field $\phi$ with a standard kinetic term and a potential $V(\phi)$, and is fully characterized by the action: 
\begin{align} \label{Eq:quintessenceAction}
S_\phi \equiv \int d^4x \, \sqrt{-g} \left( \frac12 \nabla_\mu \phi \nabla^\mu \phi - V(\phi) \right) \,.
\end{align}
This is added to the Einstein-Hilbert action for gravity and the total matter action to give a working cosmological model.
 
The dynamical behavior of a FRW background is given by the two Friedmann equations for $\mathcal{H} \equiv d \ln a / d\tau $ in conformal time $\tau$,  and the scalar field equation of motion: 
\begin{align} \label{Eq:FRWequations}
& 3 M_P^2 \mathcal{H}^2 =  \frac{\dot{\phi}^2}{2} + V(\phi) a^2 + \rho_{\rm m} a^2 \,, \nonumber \\
& 6 M_P^2 \dot{\mathcal{H}} = -2 \dot{\phi}^2 +2 V(\phi) a^2 -(\rho_{\rm m} + 3 P_{\rm m} )a^2 \,, \nonumber \\
& \ddot{\phi} +2 \mathcal{H} \dot{\phi} + a^2 \frac{\partial V}{\partial \phi} = 0 \,,
\end{align}
where the dot represents a derivative with respect to conformal time; the subscript $\rm m$ indicates a sum over all matter species that, in total, have density $\rho_{\rm m}$ and pressure $P_{\rm m}$, and evolve according to their continuity equations; $M_P^2 = 1 / (8\pi G_N)$ denotes the Planck mass. Note that any solution satisfying the first two equations will also satisfy the third equation.

The expansion history of a given model for the potential is fully determined by the field boundary conditions $(\phi_0, \dot{\phi}_0)$ at a given time. These two boundary conditions are effectively extra parameters of quintessence models, in addition to the parameters specifying the potential.
The $\Lambda$CDM limit of quintessence is given by a constant potential, set to match the value of the cosmological constant, and field initially at rest $\dot{\phi}_0 = 0$. 

As we can see in Eq.~\eqref{Eq:FRWequations}, there is no explicit dependence on the value of the field and all dependencies go through the quintessence potential. It is then possible to shift the potential, $V(\phi) \rightarrow V(\phi -\phi_c)$, shift boundary conditions and obtain the exact same cosmology.
This makes it impossible to design a unique potential as a function of $\phi$ to give the desired expansion history and 
effectively makes it impossible to perform data reconstructions of $V(\phi)$.

This is crucially where it helps to take the approach of the Effective Field Theory of Dark Energy (EFTofDE)~\cite{Bloomfield:2012ff, Gubitosi:2012hu} which, assuming isotropy and homogeneity, parameterizes all deviations from the $\Lambda$CDM background allowed by these symmetries. 
Quintessence is in the model space covered by the EFTofDE and is described by two free functions of time, denoted $\Lambda$ and $c$, in the action:
\begin{align} \label{Eq:EFTBackgroundAction}
S_{\Lambda,c} \equiv \int d^4x \sqrt{-g}  \left[ \Lambda(\tau) - c(\tau)\,a^2\delta g^{00} \right] \,,
\end{align}
in addition to the gravity and matter actions. 
Eq.~\eqref{Eq:EFTBackgroundAction} explicitly breaks time diff invariance which is ultimately what gives the quintessence scalar field as the Stueckelberg field restoring this symmetry.
One can build the correspondence between the two actions mapping one into the other with:
\begin{align} \label{Eq:quintessenceMapping}
\Lambda =&\, \frac{1}{2a^2} \dot{\phi}^2 -V(\phi)  \,, \nonumber \\
c =&\, \frac{1}{2a^2} \dot{\phi}^2 \,.
\end{align}
Friedmann equations are then given by:
\begin{align} \label{Eq:EFTFRWequations}
3 M_P^2 \mathcal{H}^2 &= 2 c a^2 - \Lambda a^2 + \rho_{\rm m} a^2  \,, \nonumber \\
6 M_P^2 \dot{\mathcal{H}} &= -2(c+\Lambda) a^2 - (\rho_{\rm m} + 3 P_{\rm m}) a^2 \,,
\end{align}
that we can use to verify the mapping relations in Eq.~\eqref{Eq:quintessenceMapping}.
We can combine the two Eqs.~\eqref{Eq:EFTFRWequations} to obtain a differential equation for $\mathcal{H}$ as a function of $\Lambda$ only.

As discussed in~\cite{Raveri:2017qvt} the dynamics are then fully set by $\Lambda$.
Then Eqs.~\eqref{Eq:EFTFRWequations} can be used to determine $c$ so that the relationship between $\Lambda$ and $c$ is consistent with the $\phi$ equation of motion. 
Since the $V, \ \phi$ formulation and the $\Lambda, \ c$ EFT formulation of quintessence are equivalent, it is possible to compute all quantities of interest from $\Lambda$ and $c$.

We start with the equation of state of dark energy that can be computed as:
\begin{align} \label{Eq:wde}
w_{\rm DE} \equiv \frac{P_{\rm DE}}{\rho_{\rm DE}} \equiv \frac{\frac{1}{2a^2} \dot{\phi}^2 -V}{\frac{1}{2a^2} \dot{\phi}^2 + V }    = \frac{\Lambda}{2c-\Lambda} \,. 
\end{align}
In quintessence models this is bounded to be $w_{\rm DE} \geq -1$ by the requirement that kinetic energy density of the scalar field has to be positive~\cite{Vikman:2004dc, Hu:2004kh, Caldwell:2005ai, Creminelli:2008wc}.

Following~\cite{Caldwell:2005tm} we can describe quintessence models as ``thawing'' and ``freezing'' by looking at $w_{\rm DE}$ and its time derivative.
This forms the $(w_{\rm DE}, w_{\rm DE}^\prime)$ plane that has been thoroughly studied in literature, {\it i.e.}~\cite{Scherrer:2005je, Linder:2006sv, Cortes:2009kc}.
``Thawing'' refers to a scenario in which $\phi$ starts nearly stationary away from the minimum due to the Hubble friction in the early universe. 
This corresponds to $w_{\rm DE} \sim -1$ and time derivatives $w_{\rm DE}' \sim 0$, where $' = d /d\ln a$. 
Then, $\phi$ starts to slowly roll down towards the minimum, which lifts $w_{\rm DE}$ above $-1$ with positive time derivatives. 
``Freezing'' is the scenario in which the opposite happens: $\phi$ starts already rolling towards the minimum, corresponding to $w_{\rm DE} >-1$. As $\phi$ approaches the minimum, the potential and the Hubble friction slow down the field velocity and $w_{\rm DE}$ approaches -1, settling there asymptotically in the future. 

The quintessence potential, as a function of time, can be computed as:
\begin{align} \label{Eq:Va}
V(a) = c - \Lambda \,. 
\end{align}
This quantity is of special interest for building working models of quintessence and, more recently, due to the Swampland Conjectures~\cite{Ooguri:2018wrx, Agrawal:2018own} putting bounds on properties of the potential.
Given the difficulties with constructing a positive cosmological constant from the theory space of string theory, recent literature has conjectured that dark energy must be a scalar field that undergoes non-trivial changes throughout cosmic history such that it is decidedly unlike a cosmological constant. These conjectures state that for a single scalar field, in Planck mass units,
\begin{align} \label{Eq:SwamplandConj}
\abs{\nabla_\phi V}/V \gtrsim \mathcal{O}(1) \quad &\textbf{or} \quad -\nabla_\phi^2 V/V \gtrsim \mathcal{O}(1)  \nonumber \\
\abs{\Delta\phi} &\lesssim \mathcal{O}(1) 
\end{align}
The ratio of the absolute value of the gradient of the potential and its value must be bound from below by a  number of order unity. 
Or, the ratio of the second derivative and the potential value must be bound from above by a negative number of order unity. These two conditions rule out behavior that mimics a cosmological constant. 
In addition, field traversal, $\abs{\Delta\phi}$, must be bounded from above by a number of order unity. 
For a discussion of the cosmological implications of the Swampland Conjectures see e.g.~\cite{Kinney:2018nny, Raveri:2018ddi}.
Both quantities in Eqs.~\eqref{Eq:SwamplandConj} can be computed in terms of $\Lambda$. 
Suppose that we have $\Lambda$ as functions of e-folds, $N = \ln a$, and we can compute $c$ and their derivatives. Then we can also derive $V$ as a function of e-folds. The quantities of interest evolve as follows: 
\begin{align} \label{eq:phidir}
\frac{d\phi}{d N} = \left(\frac{dN}{d\tau}\right)^{-1} \frac{d\phi}{d\tau}= \frac{\sqrt{2 c a^2}}{\mathcal{H}} \,,
\end{align}
where we assumed that $\phi$ increases monotonically at least at times of interest.
With these we can compute the quantities relevant to the Swampland Conjectures:
\begin{align}
\frac{\nabla_\phi V}{V} &= \frac1V \frac{dV/dN}{d\phi/dN} = \frac{d \ln V}{dN} \sqrt{\frac{\mathcal{H}^2}{2ca^2}} \,, \nonumber \\
\frac{\nabla_\phi^2 V}{V} &= \frac1V \left( \frac{\partial_N^2 V}{(\partial_N \phi)^2} - \frac{(\partial_N V)( \partial_N^2\phi ) }{(\partial_N\phi)^3}  \right) \,, \nonumber \\
\Delta\phi &= \int d\phi = \int_{N_0}^N \frac{\sqrt{2ca^2}}{\mathcal{H}}  \, dN \,,
\end{align}
where $N_0$ is the cosmic time with respect to which one chooses to compute field traversal. 
Note that these equations highlight that the value of the scalar field has no physical meaning; the  quantity  relevant for cosmology is its variation.

The cosmological dynamics of quintessence models have often been studied using a set of phase space variables~\cite{Copeland:1997et} that describe the field kinetic and potential energy.
These were used to study the presence of attractor solutions that mitigate the need to fine-tune quintessence models.
Starting from $\Lambda, \ c$ we can compute these as well:
\begin{align} \label{eq:dym}
x &= \frac{\dot{\phi}}{\sqrt{6} \mathcal{H} } = \frac{1}{\sqrt{6}} \frac{d\phi}{dN} =  \sqrt{\frac{ca^2}{3\mathcal{H}^2}} \,,\\ 
y &= \frac{\sqrt{V} a}{\sqrt{3} \mathcal{H} } = \sqrt{\frac{(c-\Lambda)a^2}{3\mathcal{H}^2}} \,,
\end{align}
and notice that these variables are defined such that:
\begin{align}
x^2 + y^2 = \frac{\frac12 \dot{\phi}^2 + V a^2}{3 \mathcal{H}^2} = \Omega_{\rm DE} \leq 1 \,,
\end{align}
where $x^2$ and $y^2$ represent the contributions of the scalar field kinetic and potential energy respectively.

It is impractical to consider the bare value of $\Lambda$ so in the following we consider relative variations with respect to the cosmological constant value of the corresponding $\Lambda$CDM model and define:
\begin{align}
\frac{\Delta\Lambda}{\Lambda} \equiv \frac{\Lambda(a)}{3(1-\Omega_{\rm m})M_P^2\mathcal{H}_0^2} -1  \,.
\end{align}

\section{Datasets and methodology} \label{Sec:Methods}

To perform the EFT reconstruction we follow the  methodology outlined in~\cite{Raveri:2019mxg}.
This follows from~\cite{Crittenden:2011aa} and allows one to reconstruct a function of time by only imposing a correlation prior on its time variations. 
This type of prior enforces smoothness of the reconstructed function and behaves as a low-pass filter that allows slow variations in the reconstructed functions to go through while penalizing fast variations.
Similar reconstruction approaches have already been successfully applied to different phenomenological DE/MG properties~\cite{Huterer:1998qv,Saini:1999ba,Huterer:2000mj,Chiba:2000im,Tegmark:2001zc,Huterer:2002hy,Huterer:2006mv,Sahni:2006pa,Mortonson:2008qy,Mortonson:2009hk,Zhao:2012aw,Vanderveld:2012ec,Wang:2015wga,Zhao:2017cud,Miranda:2017mnw,Li:2018nlh,Wang:2018fng,Li:2018dcf}.

Following~\cite{Raveri:2017qvt} we impose a minimum correlation length of $\Delta a = 0.3$ and we reconstruct the EFT function $\Delta \Lambda(a) / \Lambda$ in the range $a\in [0.1, 1.0]$, imposing  the early times limit that the $\Lambda$CDM scenario be recovered.
This is ultimately due to the fact that the theoretical prior for the correlation length that we use was derived in~\cite{Raveri:2017qvt} only at late times. 
This leaves out early dark energy type models~\cite{Poulin:2018cxd, Lin:2019qug} that we will discuss in a separate paper.

Once the reconstruction range and the prior correlation length are fixed we can represent the EFT function of interest as a spline going through a sufficient number of nodes per correlation length. 
In our reconstruction range we have three full correlation lengths and we pick 15 spline nodes. 
This ensures that we have five spline nodes per correlation length. Adding more nodes would not change results since sub-correlation length variations are suppressed by the correlation prior.

The assumption of a $\Lambda$CDM cosmology at early times influences the first correlation length that spans redshifts greater than $z\sim 1.5$.
For this reason we exclude the region $z > 1.5$ when reporting quantitative results.
In plots we will highlight with a gray shading the part that is influenced by this constraint.
The correlation prior decays exponentially so that one correlation length from the $\Lambda$CDM bound is enough for its impact to be negligible at later times.

To perform the quintessence reconstruction we use three different combinations of datasets.
We start with a ``baseline'' dataset that includes:
the Planck 2018 measurements of CMB temperature and polarization at small (Planck 18 TTTEEE) and large angular scales (lowl+lowE)~\citep{Aghanim:2018eyx, Aghanim:2019ame}; the CMB lensing potential power spectrum in the multipole range $40 \leq \ell \leq 400$~\citep{Aghanim:2018oex}; BAO measurements from BOSS DR12~\cite{Alam:2016hwk}, SDSS Main Galaxy Sample~\cite{Ross:2014qpa}, and 6dFGS~\cite{Beutler:2011hx}.
This dataset provides our baseline constraining power on all $\Lambda$CDM cosmological parameters and low redshift distances since geometric degeneracies from the CMB alone are broken.

\begin{figure*}[!ht]
\centering
\includegraphics[width=\textwidth]{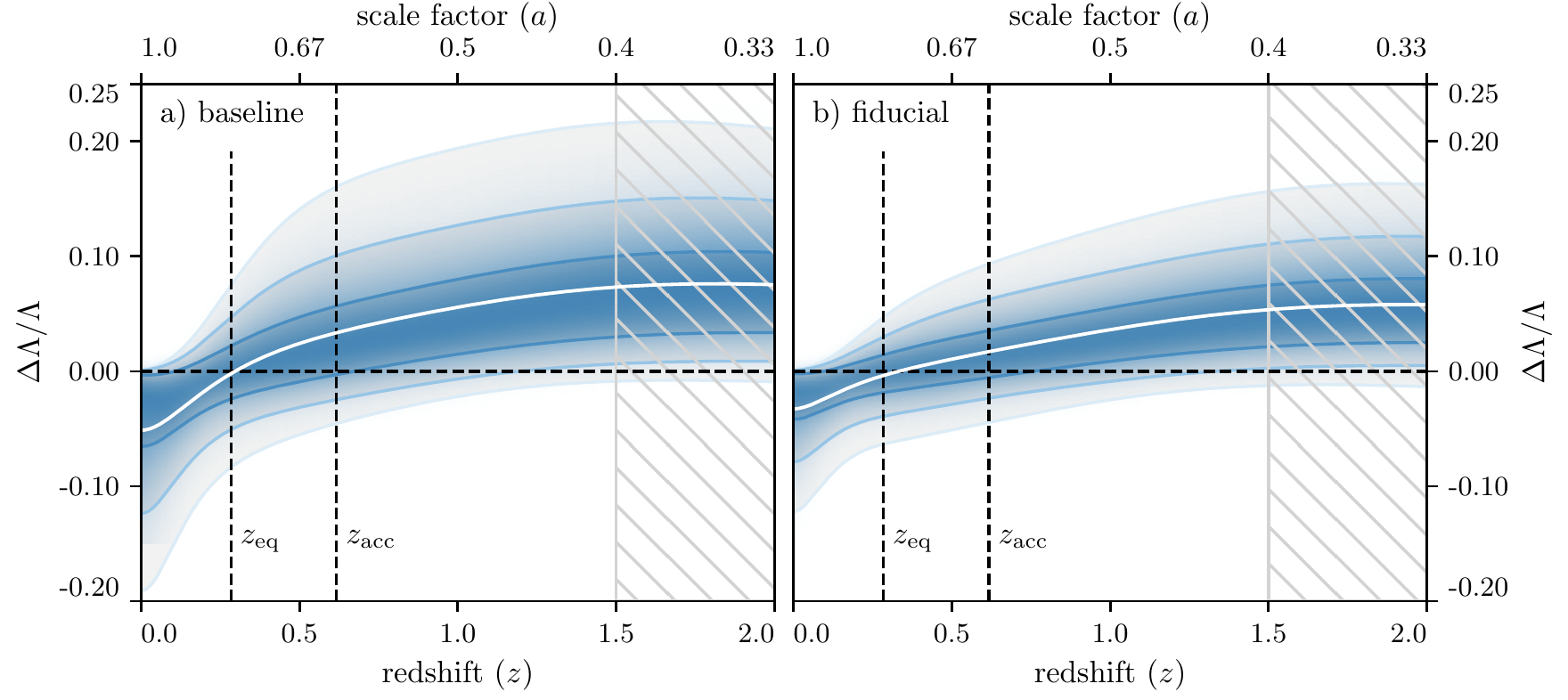}
\caption{ \label{fig:lambda}
The data constraints on $\Delta \Lambda / \Lambda$ as a function of redshift and scale factor.
Panel~a) shows the reconstruction obtained by fitting our baseline dataset combination of CMB and BAO measurements; 
Panel~b) shows the reconstruction with the addition of SN and DES data.
In both panels the white line shows the mean reconstructed behavior, solid lines from darkest to lightest show the $68\%$, $95\%$, and $99.7\%$ C.L. regions.
The two dashed vertical lines show the redshift of dark matter/dark energy equality, $z_{\rm eq}$, and the redshift when cosmic acceleration begins, $z_{\rm acc}$.
The horizontal dashed line shows the $\Lambda$CDM limit of the model.
The grey shading highlights the first reconstruction correlation length that is influenced by the assumption that the $\Lambda$CDM model should be recovered at early times.
}
\end{figure*}

In addition to this we consider the Pantheon Supernovae (SN) sample~\cite{Scolnic:2017caz} that provides relative distance measurements in the redshift range $z\in[0.01,2.26]$.
Then we add measurements of the clustering and lensing of large scale cosmological structures from the Dark Energy Survey (DES) via its Year 1 dataset~\cite{Abbott:2017wau}.
We consider the DES measurements of galaxy clustering, lensing and their cross correlation ($3\times 2$) at large angular scales that are independent of the modeling of non-linear clustering~\cite{Abbott:2018xao}.
This ``fiducial'' data set thus includes all datasets in the baseline combination along with SN and DES.

We investigate the impact of the Hubble constant tension on the quintessence reconstruction using the SH0ES measurement of the Hubble constant, $H_0=74.03\pm1.42$ (in units of km\,s$^{-1}$\,Mpc$^{-1}$ here and throughout)~\citep{Riess:2019cxk}.
Since late-times quintessence models cannot relieve this tension~\cite{Raveri:2018ddi, Benevento:2020fev} we do not include the Hubble constant measurement in the fiducial dataset combination that we use. 
Further we note that there are other measurements of the Hubble constant that have lower values~\cite{Freedman:2019jwv, Riess:2020fzl} but we do not consider them since they would not alter the results qualitatively.

To summarize, we follow the naming convention:
\begin{itemize}
\item Baseline = Planck 18 TTTEEE + lowl + lowE + CMB lensing + BOSS DR12 BAO + SDSS MGS BAO + 6dFGS BAO
\item Fiducial = Baseline  + Pantheon SN + DES Y1 $3\times 2$
\item Fiducial + $H_0$
\end{itemize}

In addition to the parameters used for the quintessence reconstruction we include the standard 6 parameters of the
$\Lambda$CDM model: 
baryon density $\Omega_b h^2$;
cold dark matter density $\Omega_c h^2$; 
the initial curvature spectrum normalization at $k=0.05$ Mpc$^{-1}$ $A_s$, and its tilt $n_s$;
the optical depth to reionization $\tau$;
the angular size of the CMB sound horizon $\theta_s$.
All these parameters have the usual non-informative priors~\cite{Aghanim:2019ame}.
We include all the recommended parameters and priors describing systematic effects in the different data sets we consider.
We fix the sum of neutrino masses to the minimal value~(e.g.~\cite{Long:2017dru}).

To produce cosmological predictions and compare them to data, we use the EFTCAMB and EFTCosmoMC codes~\cite{Hu:2013twa,Raveri:2014cka}, modifications to the Einstein-Boltzmann code CAMB~\cite{Lewis:1999bs} and the Markov Chain Monte Carlo (MCMC) code CosmoMC~\cite{Lewis:2002ah} respectively.
For the statistical analysis of the MCMC results we use GetDist~\cite{Lewis:2019xzd}.
We sample the posterior parameter distribution until the Gelman-Rubin convergence statistic~\cite{gelman1992} satisfies $R-1<0.01$ or better.

\section{Results} \label{Sec:Results}
\subsection{Reconstruction of quintessence}

We start with constraints on $\Delta \Lambda / \Lambda$ as it is the quantity that is directly estimated from the data.
Other results are derived from the $\Delta \Lambda / \Lambda$ reconstruction along with the standard  cosmological parameters.

In Fig.~\ref{fig:lambda} we show $\Delta \Lambda / \Lambda$ considering the baseline and fiducial dataset combinations.
As we can see in Panel a) the baseline data combination constrains $|\Delta \Lambda / \Lambda| < 15\%$ at $95\%$ C.L. at all times of interest.
The constraint that the present day value of $\Delta \Lambda / \Lambda$ should be negative follows from the positive definiteness of the scalar field kinetic energy density. At earlier times there is no such constraint.
By comparing the two panels of Fig.~\ref{fig:lambda} we can see that the fiducial data combination is sensibly more constraining, enforcing $|\Delta \Lambda / \Lambda| < 11\%$ at $95\%$ C.L. at all redshifts of interest.
This improvement is mostly driven by SN data that span a large redshift range with accurate distance measurements.
From the figure we do not see any significant deviation from the $\Lambda$CDM model, and have   confirmed that the $\chi^2$ of the fit does not show  significant improvement over $\Lambda$CDM  for either dataset combination.
We note that the mean seems to show deviations from the $\Lambda$CDM model because of non-Gaussianity in the marginalized $\Delta \Lambda / \Lambda$  posterior close to the $\Lambda$CDM limit.

We can compute how many quintessence parameters are constrained by data over the prior, as discussed in~\cite{Raveri:2018wln}, by comparing the prior and posterior covariances of the reconstruction parameters, {\it i.e.} the values of $\Delta \Lambda / \Lambda$ at the spline nodes.
We find that our baseline data combination constrains four parameters, in addition to the six $\Lambda$CDM parameters.
Our fiducial dataset constrains five.
We notice that this number is higher than what reported in~\cite{Raveri:2019mxg} because of the updated datasets we use, including the latest release of CMB measurements from Planck and galaxy clustering from DES.
This number is also significantly higher than what is commonly used in phenomenological parametrizations of the DE equation of state that are generally limited to one or two extra parameters. 
This means that these simpler parametrizations, while they might still allow the detection of deviations from $\Lambda$CDM, are not guaranteed to allow the data to express their full constraining power.

\begin{figure}[!th]
\centering
\includegraphics[width = \linewidth]{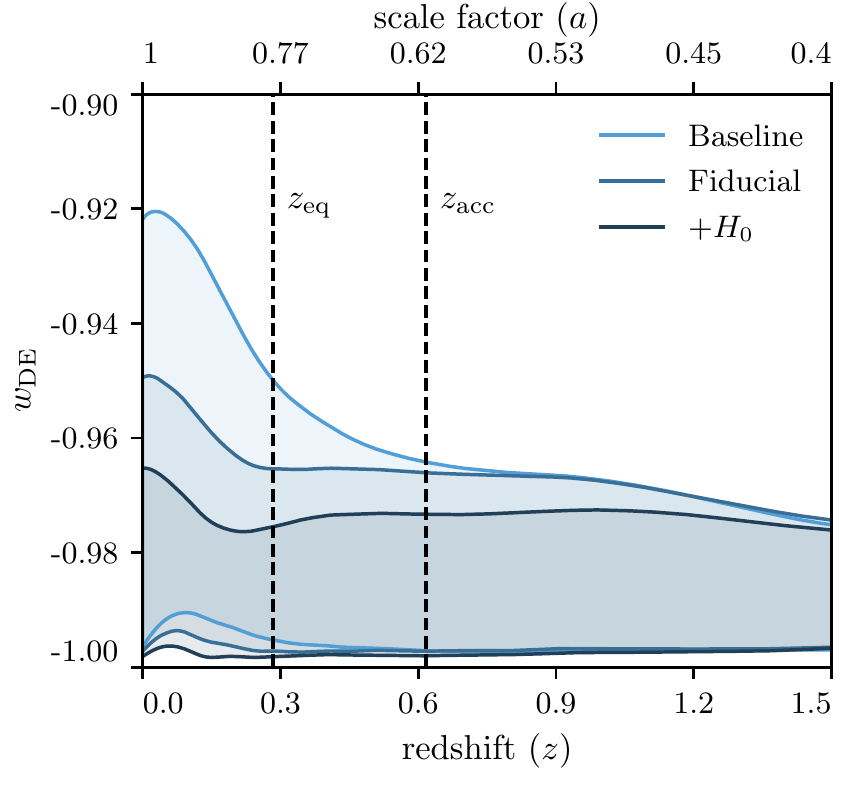}
\caption{ \label{fig:wde}
Constraints from data on the dark energy equation of state, $w_{\rm DE}(z)$, as a function of redshift.
The colored bands show the 68\% C.L. regions, as a function of time, for the different dataset combinations we consider, as shown in legend.
Other figure conventions follow Fig~\ref{fig:lambda}.
We note that the model has to have $w_{\rm DE}(z) \geq -1$ to satisfy the theoretical viability requirement of positive kinetic energy density.
}
\end{figure}

We can propagate the constraints on $\Delta \Lambda / \Lambda$ to constraints on $w_{\rm DE}(z)$, as discussed in Sec.~\ref{Sec:Methods}.
The results are shown in Fig.~\ref{fig:wde} where we can see the 68\% C.L. constraints on $w_{\rm DE}(z)$ from the different dataset combinations we consider.

As we can see for all dataset combinations $w_{\rm DE}(z)$ is forced against the $\Lambda$CDM limit.
This limit cannot be crossed since quintessence models require that $w_{\rm DE}(z) \geq -1$ at all times.
The baseline data combination pushes toward that boundary since CMB+BAO slightly prefers $w_{\rm DE}(z) < -1$, as noted in~\cite{Aghanim:2018eyx}.
Additional datasets do not change this appreciably but are more constraining, especially at late times.
In particular a large increase in constraining power results from the addition of the $H_0$ measurement.
This reflects the fact that quintessence cannot solve the tension between local and CMB distance calibrations as that would require $w_{\rm DE}<-1$.
So the $H_0$ measurement appears to provide more constraining power than is reasonable, which is the reason  we exclude it from the fiducial results.

At all times the DE equation of state is constrained to be $0 \leq 1+w_{\rm DE} \leq 0.15$ at 95\% C.L. by the baseline dataset and $0 \leq 1+w_{\rm DE} \leq 0.10$ by the fiducial dataset, compatible with the bounds on $\Delta \Lambda / \Lambda$.

\begin{figure*}[!ht]
\centering
\includegraphics[width=\textwidth]{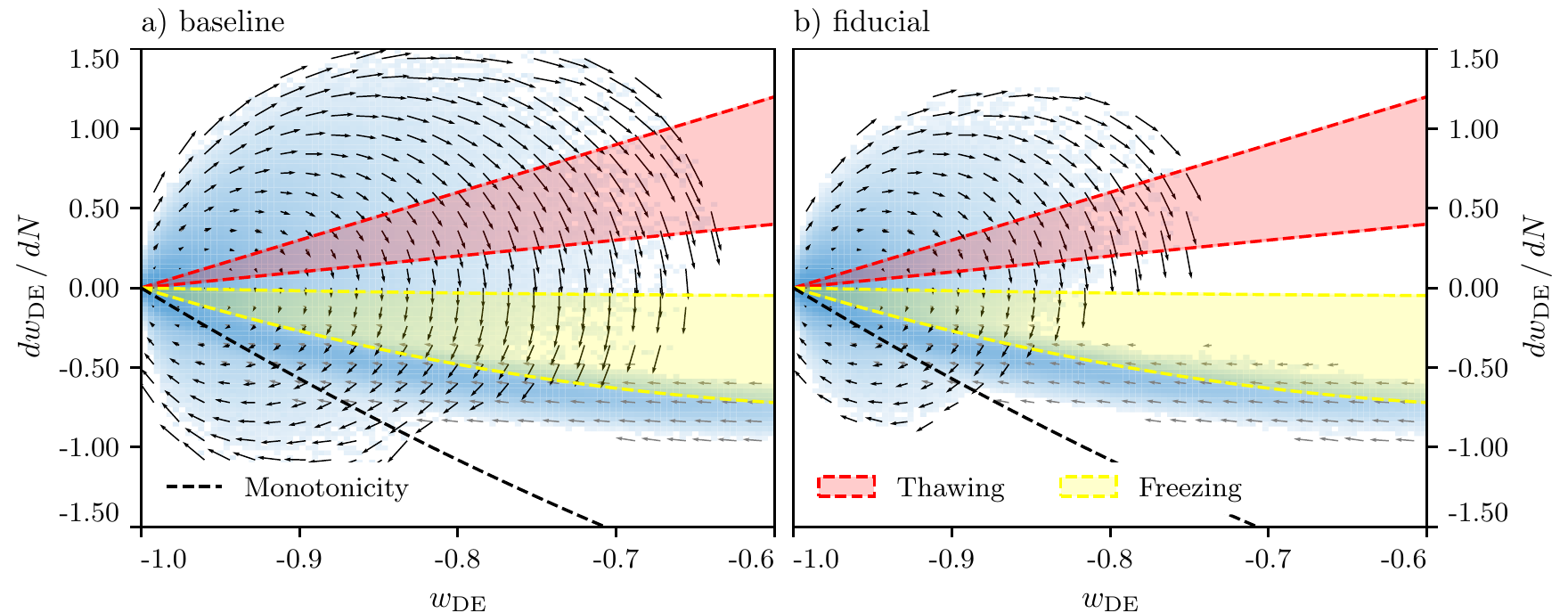}
\caption{ \label{fig:thaw}
Reconstructed trajectories in the $(w_{\rm DE}, w_{\rm DE}^\prime)$ plane for two different dataset combinations.
In both panels the shading represents the number density of trajectories.  The arrows indicate the average velocity $(w_{\rm DE}^\prime, w_{\rm DE}^{\prime\prime})$ at a given position.
The gray arrows highlight the behavior of the reconstructed trajectories before data constraints enter at $z=1.5$.
The red lines enclose the ``thawing'' region from~\cite{Caldwell:2005tm} and the yellow lines enclose the ``freezing'' region. 
The black dashed line line shows where $w_{\rm DE}^\prime = -3(1-w_{\rm DE}^2)$~\cite{Scherrer:2005je}.
Instead of a clear ``freezing'' or ``thawing'' behavior,  trajectories reconstructed from the data start in the ``freezing'' region, reach $w_{\rm DE}=-1$, and oscillate around that point as a consequence of the data's preference for $w_{\rm DE} \approx -1$.
}
\end{figure*}

\subsection{$(w_{\rm DE}, w_{\rm DE}^\prime)$ phase space trajectories} \label{Sec:wwp}

From the DE equation of state we can reconstruct trajectories in the $(w_{\rm DE}, w_{\rm DE}^\prime)$  phase space, proposed by~\cite{Caldwell:2005tm} in the context of ``freezing'' and ``thawing'' scenarios.
In Fig.~\ref{fig:thaw} we show the data constraint on the $(w_{\rm DE}, w_{\rm DE}^\prime)$ plane from the baseline and fiducial datasets.
The shading represents the hit-count of how many times the $(w_{\rm DE}, w_{\rm DE}^\prime)$ trajectories passed through a certain bin of the phase space. 
The arrows represent the average velocity vector of the $(w_{\rm DE}, w_{\rm DE}^\prime)$ trajectories at a given bin.
Gray arrows show the behavior of trajectories in the regime where there is no data constraint ($z>1.5$) while the black arrows highlight behavior in the data constrained regime ($z\in [0,1.5]$).
The red region is approximately the allowed region for ``thawing'' scenarios and the yellow for ``freezing'' scenarios as described by~\cite{Caldwell:2005tm}.

These plots show that in general the reconstructed cosmologies do not appear to follow either the ``freezing'' or ``thawing'' scenarios. 
This is not to say that we can rule out those scenarios. The constraining power necessary for such a distinction is on the order of $\sigma(w^\prime_{\rm DE} ) \approx 1+w_{\rm DE}$~\cite{Scherrer:2005je, Caldwell:2005tm}.
For $z<1.5$ where we actually have data constraints, $1+w_{\rm DE} \lesssim \mathcal{O}(0.01)$ and $\sigma(w^\prime_{\rm DE})\sim \mathcal{O}(0.1)$. That is, the constraining power we have on $w_{\rm DE}^\prime$ is about 1 order of magnitude too weak to rule out either scenarios.
Rather, trajectories come in toward $(-1,0)$ from the ``freezing'' region and once that point is reached start circling around that point.
In doing so, some trajectories violate the $w_{\rm DE}^\prime > -3(1-w_{\rm DE}^2)$ limit set in~\cite{Scherrer:2005je}. That limit assumes the potential to be monotonically decreasing in $\phi$ whereas our results do not~\cite{Huterer:2006mv}. A sizable portion of the trajectories have non-monotonic potentials, as will be discussed later in relation to the Swampland Conjectures.

Fig.~\ref{fig:thaw} highlights one aspect of the DE equation of state reconstruction that was hidden in Fig.~\ref{fig:wde}.
Oscillations in $w_{\rm DE}$ weaken integral constraints. 
Present constraints on $w_{\rm DE}$ come from distance measurements which are related to integrals of $w_{\rm DE}$ instead of its actual value at different redshifts. The phase and frequency of oscillations in $w_{\rm DE}$ are then weakly constrained by data. This is similar in nature to the argument given in~\cite{Maor:2000jy}.
Since the data prefers $w_{\rm DE}=-1$ and since physical viability enforces $w_{\rm DE} \geq -1$ the reconstructed $w_{\rm DE}(z)$ oscillates close to the $w_{\rm DE}=-1$ line. 
In Fig.~\ref{fig:wde} these oscillatory modes in the $w_{\rm DE}$ reconstruction are averaged away and the 68\% C.L. bounds look non-oscillatory.
The oscillations are  not synchronized across samples, so averaged over many samples the oscillatory behavior is not visible.
In the $(w_{\rm DE}, w_{\rm DE}^\prime)$ space these bounces clearly look like circles as in the phase space of the harmonic oscillator.
As we add more data constraints, in going from Panel a) to Panel b), we shrink the amplitude of these oscillations and hence the viable volume in phase space.
This shows that, in order to distinguish between ``thawing'' and ``freezing'' models, independent of assumptions about the potential, we need to break the intrinsic data degeneracy that allows for oscillatory modes in $w_{\rm DE}(z)$.

\begin{figure}[!th]
\centering
\includegraphics{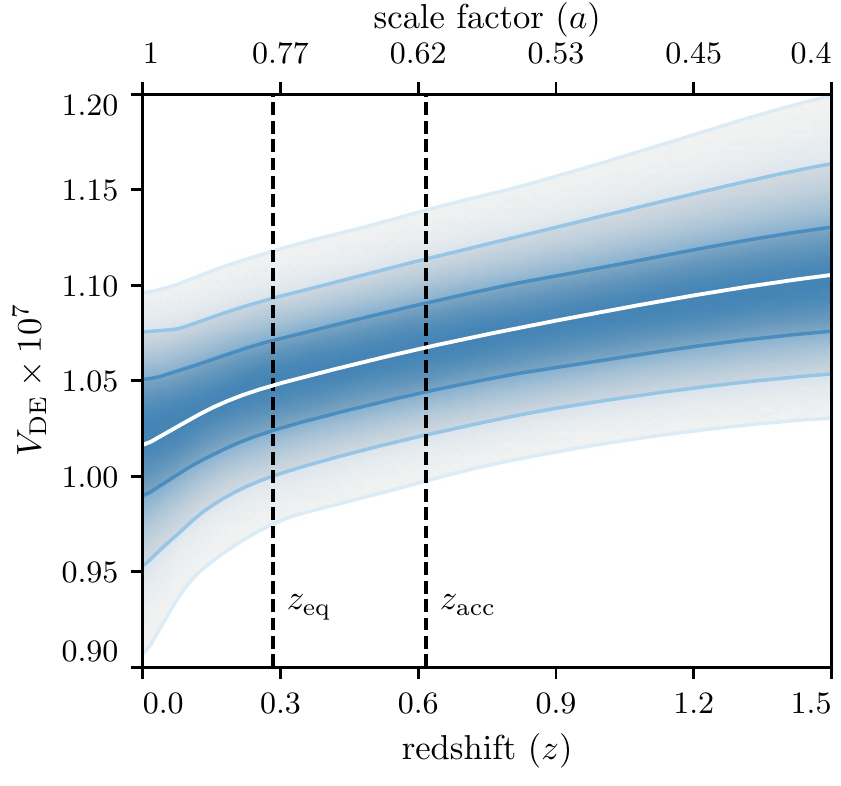}
\caption{ \label{fig:vde}
The reconstructed quintessence potential, $V(z)$, as a function of redshift for the fiducial dataset combination. Conventions for the different confidence interval curves follow Fig~\ref{fig:lambda}.  Though the $V(z)$ behavior generally resembles a downhill motion of $\phi$, we note that individual $V(z)$ trajectories are not monotonic, as was the case with $w_{\rm DE}$.
}
\end{figure}

\subsection{The Swampland Conjectures and the quintessence potential}

We now turn to quintessence physical properties rather than phenomenological quantities.
In Fig.~\ref{fig:vde} we show the reconstruction of the quintessence potential $V(z)$ as a function of redshift with the fiducial dataset combination. 
Other datasets do not qualitatively alter this picture but only quantitatively impact the error bars.
As we can see the overall potential scale is set by the required energy density of 
DE at late times. At earlier times the potential is increasing as the field had to roll downhill.
We note that, similarly to what happens for the equation of state, while the averaged reconstruction looks monotonic and smooth the single potential samples are usually not.

\begin{figure}[!th]
\centering
\includegraphics{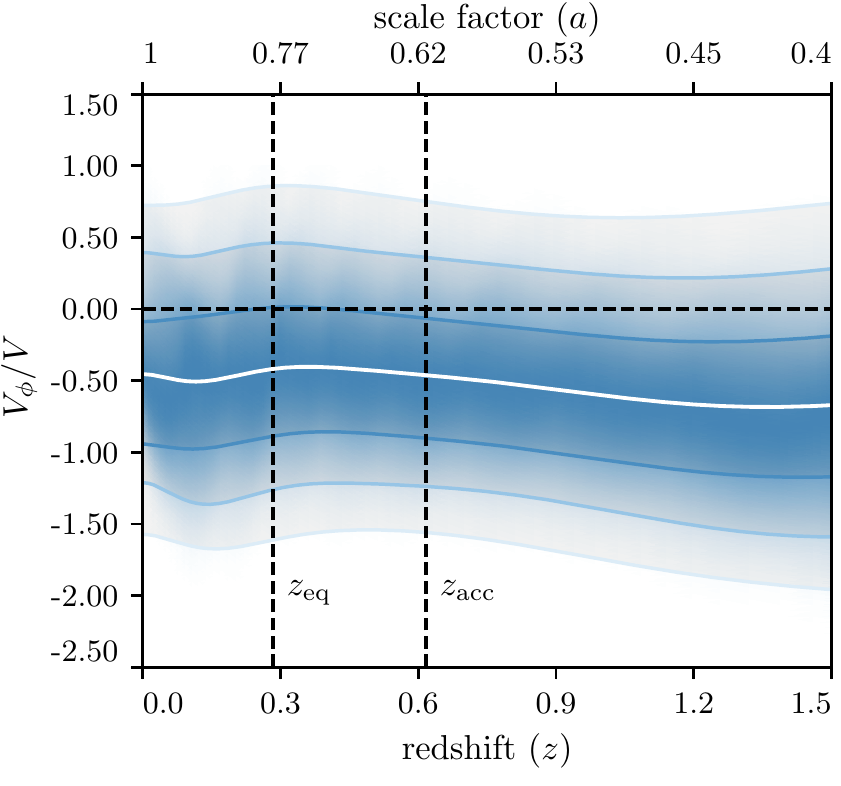}
\caption{ \label{fig:vpv}
Reconstruction of the derivative of the quintessence potential, $V_\phi/V$, as a function of redshift for the fiducial dataset combination.
The curves showing confidence intervals   follow the conventions of Fig~\ref{fig:lambda}.
We do not find a $\mathcal{O}(1)$ lower bound to $\abs{V_\phi/V}$ and would need more constraining power to find a $\mathcal{O}(0.1)$ one.
}
\end{figure}

We can now consider the derivatives of the potential that are related to Swampland Conjectures. 
We start with the part of the conjecture that concerns the first derivative of the potential, shown in Fig.~\ref{fig:vpv}, {\it i.e.} the  reconstruction of $V_\phi/V$ from the fiducial data set. 
With present constraining power, we find at the 68\% confidence level
that $-1 \lesssim V_\phi/V \lesssim 0$ across $z\in [0,1.5]$. 
Consistent with prior work~\cite{Raveri:2018ddi}, there is
no $\mathcal{O}(1)$ lower bound to  $\abs{V_\phi/V}$. We need more
constraining power to obtain a possible upper bound of
$\mathcal{O}(0.1)$.

We have tested the second derivative of the potential but found that the constraints are weak.
This is understandable since $w_{\rm DE}$ is constrained by distance measurements. Distances are given by integrals of $w_{\rm DE}$. $w_{\rm DE}$ and $V_{\rm DE}$ are related algebraically, therefore $V_{\phi\phi}$ enters at the level of the third derivative of measured quantities. 
In a polynomial expansion third derivatives are related to the fourth coefficient so our weak constraining power on the second derivative of the potential is a reflection of the fact that present datasets measure 4 to 5 extra dark energy parameters --  constraints on the last two parameters are still  weak.

We can combine the two potential constraints in Eq.~\eqref{Eq:SwamplandConj} to see that one scenario unambiguously forbidden by this Swampland Conjecture is a local minimum in $V(\phi)$ at $V >0$ with  $\abs{V^\prime/V} = 0$ and $V^{\prime\prime}/V > 0$. We find that the $V(\phi)$ implied by about 66\% of our MCMC samples violate the conjecture as they possess a local minimum for $z\in [ 0,1.5]$. This number allows us to conclude with more concreteness that our results do not clearly favor either compliance or non-compliance with the Swampland potential constraints.

\begin{figure}[!t]
\centering
\includegraphics{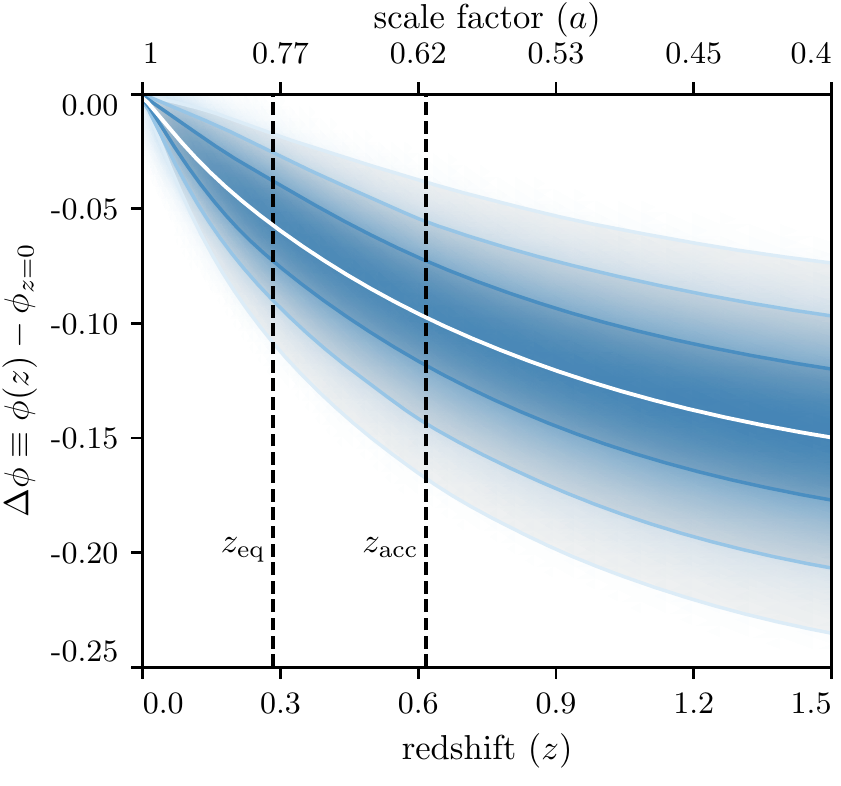}
\caption{ \label{fig:phiat}
The data reconstructed scalar field excursion, $\Delta \phi(z) = \phi(z) - \phi(z=0)$, as a function of redshift for the fiducial dataset combination.
Figure conventions follow Fig.~\ref{fig:lambda}.
To $95\%$ C.L., we find $\abs{\Delta\phi} < 0.2$, between $z=0$ and $1.5$, compatible with the Swampland Conjectures. This result is qualitatively robust to changing dataset combination.
}
\end{figure}

We then consider the second part of the Swampland Conjectures which imposes an order unity upper bound on field traversal, $\Delta \phi \lesssim \mathcal{O}(1)$.
This is tested in Fig.~\ref{fig:phiat} that computes field excursion from the present day value of the field.
With current data, the constraints are already of  order $0.1$, so we can make a more definite statements about the $\Delta \phi$ Swampland Conjecture than we can about the potential Swampland Conjecture.   
We find that at all redshifts of interest $|\Delta\phi| < 0.23$ at 95\% C.L. for our baseline dataset combination and $|\Delta\phi| < 0.20$ at 95\% C.L. for the fiducial data set, both of which are bounds compatible with the Swampland Conjectures.

\begin{figure}[!th]
\centering
\includegraphics[width=\linewidth]{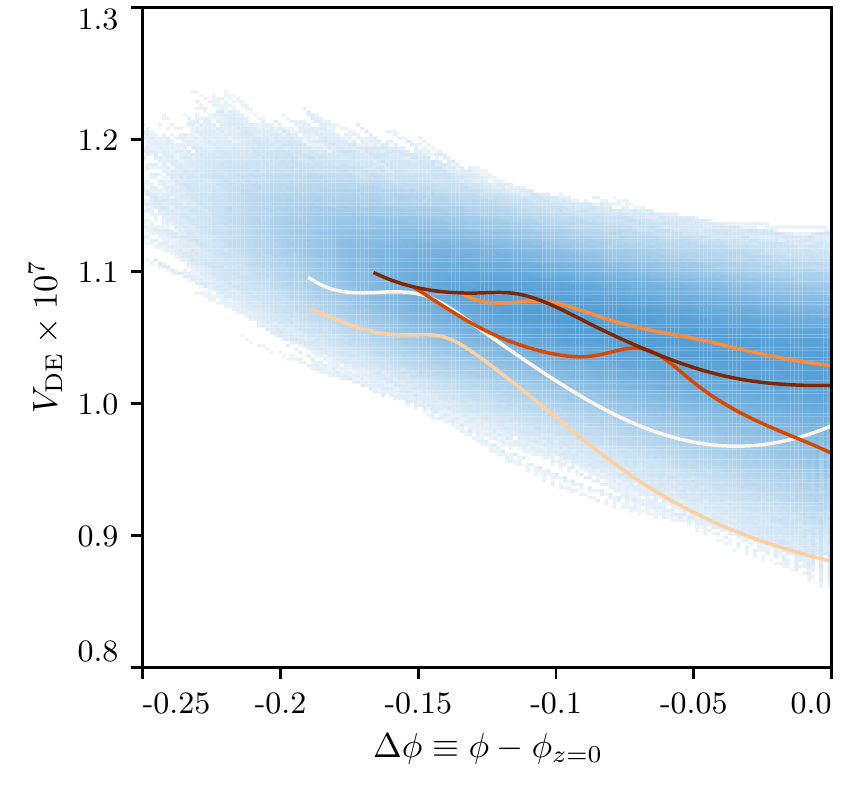}
\caption{ \label{fig:vphi}
$V(\Delta \phi)$ reconstructed for the fiducial dataset combination. 
Shading  represents the number density of $(\phi,V)$ trajectories and in addition five randomly selected reconstructions are shown by the curves. With current data, we cannot rule out non-monotonicity of $V(\phi)$. 
Note also that larger values of field excursion are correlated with larger values of the potential.
}
\end{figure}

With knowledge of $\Delta \phi(z)$ as shown in Fig.~\ref{fig:phiat} and $V(\phi(z))$ as shown in Fig.~\ref{fig:vde}, it is possible to reconstruct $V(\Delta \phi)$ without reference to $z$. The results are shown in Fig.~\ref{fig:vphi}. 
Note that, as a consequence of translation invariance of the potential, cosmological observations cannot constrain $V(\phi)$ but only $V(\Delta\phi)$, {\it i.e.} the part of the potential that is traversed during cosmological evolution.
The $(\phi, V)$ plot is a useful tool for examining general characteristics of the potential. The potential favored by the data appears to have, on average, a negative first derivative and a positive second derivative throughout $z\in [0,1.5]$. However, as Fig.~\ref{fig:thaw} hints, 69\% of our MCMC samples imply at least one extrema in their $V(\phi)$ reconstructions. Fig.~\ref{fig:vphi} shows five randomly selected samples, which typically have bumps and plateaus -- though at different values of $ \Delta \phi$. With current datasets we cannot rule out small bumps in $V(\phi)$ though the average behavior seems monotonic.

\begin{figure}[!th]
\centering
\includegraphics{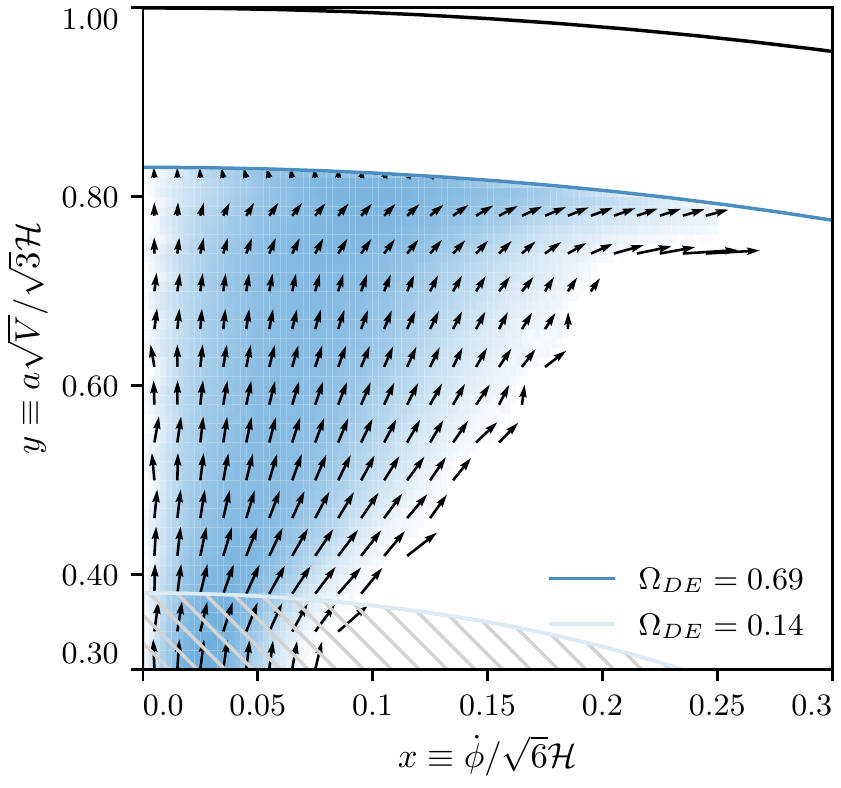}
\caption{ \label{fig:dym}
The quintessence phase space as reconstructed from our fiducial dataset combination.
The shading represents number density of trajectories and arrows the average velocity ($dx/da, dy/da$) at a given position in phase space.
The gray shaded area indicates the region with of no data constraints. 
Since  $x^2+y^2 = \Omega_{\rm DE}$ different solid circles indicate relevant values of $\Omega_{\rm DE}$ as shown in legend.
The general flow in phase space is upwards, compatible with increasing DE density as a function of time.
Using k-means clustering we find no definitive attractor, as explained in the text.
}
\end{figure} 

\subsection{Dynamical phase space trajectories}
We can now look at the phase space of quintessence as defined in Eq.~\eqref{eq:dym}. 
This is shown in Fig.~\ref{fig:dym} for our fiducial dataset combination. 
Like in the $(w_{\rm DE}, w_{\rm DE}^\prime)$ phase space diagram, the background color indicates number density of trajectories in phase space, as allowed by data constraints, and the arrows are the average velocity in phase space at a given position. 
Since $x^2+y^2 = \Omega_{\rm DE}$, different solid lines indicate its early time value at $z=1.5$ and today. The gray box is the region where trajectories are not constrained by data.
As we can see the phase space of quintessence is heavily constrained by data and there is no clear attractor to which the trajectories seem headed. 
Using the k-means clustering algorithm to detect clusters of paths and their representative dominant paths in the phase space, we find that the dominant paths' end points are scattered across the $\Omega_{\rm DE}=0.69$ line as one may expect from the arrows. This means that the data constraints do not appear to favor an attractor solution.
The general flow is upwards and only slightly to the right, meaning that it is mostly the potential energy of $\phi$ that comes to dominate the universe. The conclusions are mostly the same with other dataset combinations. 
More constraining combinations just reduce the viable volume of phase space and does not change conclusions qualitatively.

\section{Conclusions} \label{Sec:Conclusions}
Using  state of the art cosmological observations we have performed a fully non-parametric EFT based reconstruction of quintessence models from data. In particular, we make no assumptions about the potential of the scalar field. 
We show that our methodology and current data constraints are robust enough to study various theoretical aspects of quintessence. 
The datasets used are the CMB measurements from the Planck satellite, BAO measurements from BOSS, measurements of supernovae distances, and galaxy clustering and lensing from DES. 

The reconstructed  equation of state of DE, $w_{\rm DE}$, does not deviate from $-1$ at late times by more than $10\%$, though it is given enough degrees of freedom to do so. 
Specifically for our fiducial choice of datasets we show that in the redshift range $z<1.5$, at 95\% C.L. $0 \leq 1+w_{\rm DE} \leq 0.10$, with the $w_{\rm DE}>-1$ constraint coming from the theoretical prior for single field quintessence models.
Furthermore, reconstructing the ($w_{\rm DE} , w_{\rm DE}^\prime$) phase space, in Fig~\ref{fig:thaw}, we find that present cosmological data does not simply follow the classification in ``freezing'' or ``thawing'' scenarios. Instead, the preferred evolution of $w_{\rm DE}$ is such that it oscillates with the minimum close to $w_{\rm DE} = -1$. 
We discuss in Sec.~\ref{Sec:wwp} how the reconstructed phase space trajectories are sensitive to different assumptions in the literature about monotonicity of $V(\phi)$.
In order to identify and discriminate between ``freezing'' and ``thawing'' scenarios we find that the constraints on the time derivative of the DE equation of state will need to be an order of magnitude stronger, so such a task awaits major improvements in the available data.

From the equation of state of DE we have reconstructed the quintessence potential and obtained general constraints on Swampland Conjectures in the single field dark energy case. 
Regarding the conjecture on the slope of the potential $V(\phi)$, we find no evidence for an order unity lower bound on $\abs{V_\phi/V}$, as in Fig.~\ref{fig:vpv} -- consistent with our finding of no significant detection of deviations from $\Lambda$CDM.
On the other hand an {\it upper} bound of $\mathcal{O}(0.1)$ could be obtained in the coming years as datasets with increased constraining power will become available.
Current data constraints are also not strong enough to provide meaningful  bounds on the alternative conjecture $- V_{\phi\phi}/V \gtrsim \mathcal{O}(1)$.
Our fiducial dataset combination is, however, constraining enough to confirm the conjecture that field traversal is bound from above: we find $|\Delta\phi| < 0.20$ at 95\% C.L.

We have shown the fully reconstructed dark energy phase space, Fig.~\ref{fig:dym}, and find that the viable region is severely constrained towards the slow roll limit. 
That is, the potential energy is dominant compared to the kinetic energy of the scalar field. 
We find no indication of the existence of an attractor towards which the phase space flows are headed. 

These general conclusions apply not just for our fiducial dataset choice but are stable to the inclusion or removal of its components. More and more accurate data in the low-z, late-time regime will add constraining power, strengthening the results above. 
In particular, as shown in  Sec.~\ref{Sec:Results}, the preferred behavior of $w_{\rm DE}$ is to oscillate but the phase is entirely unconstrained. 
This occurs because present data constrain distances, {\it i.e.} integrals of $w_{\rm DE}$, at the first derivative level. 
Planned direct measurements of $H(z)$ from BAO surveys would allow us to constrain the phase of $w_{\rm DE}$ oscillations as well, and better constrain $V_{\phi\phi}/V$.
The methodology used here to constrain the EFTofDE and then translate the constraints to covariant quantities could also be used to shed light on other Swampland Conjecture-like considerations, such as those for non-canonical kinetic terms considered in~\cite{Solomon:2020viz}, as well as the forms of general Horndeski coupling functions. 

\acknowledgments
We thank 
Dragan Huterer, 
Justin Khoury, 
Eric Linder, 
Levon Pogosian 
and Paul Steinhardt 
for helpful comments.
MR is supported in part by NASA ATP Grant No. NNH17ZDA001N, and by funds provided by the Center for Particle Cosmology. 
BJ is supported in part by the US Department of Energy Grant No. DE-SC0007901.  
Computing resources are provided by the University of Chicago Research Computing Center through the Kavli Institute for Cosmological Physics at the University of Chicago. 

\bibliographystyle{apsrev4-1}
\bibliography{biblio}
\end{document}